\documentclass{ws-ijmpa}

\usepackage{amsmath}
\usepackage{amssymb}
\usepackage{overcite}

\newcommand{\half}{\frac{1}{2}}

\newcommand{\pdo}[1]{\ensuremath{\frac{\partial }
	{\partial #1 }}}

\newcommand{\slashletter}[1]{\ensuremath{\kern+0.1em /\kern-0.65em #1}}

\begin{document}

\setcounter{secnumdepth}{1}

\markboth{M. P. FRY}{Fermion Determinants}

\title{FERMION DETERMINANTS}

\author{\footnotesize M. P. FRY}

\address{School of Mathematics, University of Dublin\\
Dublin 2, Ireland}

\maketitle

\begin{abstract}
The current status of bounds on and limits of fermion
determinants in two, three and four dimensions in QED and QCD is
reviewed. A new lower bound on the two-dimensional QED determinant is
derived. An outline of the demonstration of the continuity of this
determinant at zero mass when the background magnetic field flux is
zero is also given.
\end{abstract}

\section{Introduction}

\label{Sec_Intro}

\noindent
The calculation of fermion determinants is an old problem that has
reemerged as part of mainstream physics. They lie at the heart of
gauge field theories with fermions and appear in the calculation of
every physical process. Their increasing interest is due to lattice
QCD and the improvement of lattice fermion actions. They are obtained
by integrating over the fermion fields analytically to produce the
one-loop effective action $S_{eff} = - ln \, det$, where
$det$ is formally the ratio $det(\slashletter{P} - e\slashletter{A} +
m)/det(\slashletter{P} + m)$ of Fredholm determinants of
euclidean Dirac operators. Properly defined, det is a gauge invariant
but nonlocal function of the background gauge potential $A_{\mu}$, and
it is this nonlocality that makes $det$ so challenging to calculate both
analytically and numerically. At present the best that can be hoped
for from analytic nonperturbative calculations are restrictive upper
and lower bounds on determinants in two, three and four dimensions, as
well as particular limits, such as strong coupling or small fermion
mass. Such results are relevant to lattice calculations extrapolated
to the continuum as they are a nontrivial test of lattice
discretization procedures, algorithms, and practices such as taking
the square root of the Kogut-Susskind determinant to simulate two
degenerate quark flavors.

In Sec. \ref{Sec_BoundLimitdet} we report on the current status of
bounds on and limits of fermion determinants. In
Sec. \ref{Sec_LowerBound} the derivation of a new lower bound on the
two-dimensional QED determinant is given. Section \ref{Sec_ChiralLim}
contains the proof that the chiral limit of this determinant coincides
with the Schwinger model's determinant only when the background
magnetic field's flux is zero.

\section{Bounds and Limits on det}

\label{Sec_BoundLimitdet}

\noindent
In order to make estimates the class of background gauge fields has to
be defined. Since the determinant is part of the gauge field's action,
$A_{\mu}$ and the field strength $F_{\mu\nu}$ are random
fields. However, there is a need to regulate in any dimension. For
$A_{\mu} \in S'$, the Schwartz space of tempered distributions, this
can be done by smoothing $A_{\mu}$ in the determinant and elsewhere,
except in the gauge-fixed Gaussian measure $d\mu(A)$ for $A_{\mu}$, by
convoluting it with a function $h_{\Lambda} \in S$, the functions of
rapid decrease.~\cite{Seiler75a} This gives a potential
$A^{\Lambda}_{\mu}(x) = (h_{\Lambda}*A_ {\mu})(x)$ that is a
polynomial bounded $C^{\infty}$ function whose covariance is now

\begin{equation}
\label{Eq_PotACovariance}
\int \, d\mu(A) \, A^{\Lambda}_{\mu}(x) \, A^{\Lambda}_{\nu}(y) = 
	D^{\Lambda}_{\mu\nu}(x - y).
\end{equation}

\noindent
The Fourier transform of $D^{\Lambda}_{\mu\nu}$ is proportional to
$|\hat{h}_{\Lambda}|^2$, where $\hat{h}_{\Lambda}$ is the Fourier
transform of $h_{\Lambda}$ and has the property that
$\hat{h}_{\Lambda} = 0$ for momenta $k^2 > \Lambda^2$. Therefore
without loss of generality we may assume smooth potentials and
fields. In addition they must be in some $L^{n}(\mathbb{R}^d)$
space(s) that will be stated explicitly below.

All results, except the upper bounds for $d = 2, 3$, are for an
abelian field obtained by using Schwinger's proper time
definition~\cite{Schwinger51} of the determinant as the starting
point, namely

\begin{multline}
\label{Eq_Schwingerdet}
\text{ln det} = \half \int^{\infty}_{0} \frac{dt}{t} \\
\left( \text{Tr}
	\left\{ e^{-P^2t} - \text{exp}\left[ -((D - e A)^2 +
	\half e \sigma^{\mu\nu} F_{\mu\nu})t
	\right] \right\} + 
	\frac{e^2 ||F||^2}{24 \pi^2}
\right) e^{-t m^2},
\end{multline}

\noindent
where $\sigma^{\mu\nu} = [\gamma^{\mu}, \gamma^{\nu}]/2i$,
$\gamma^{\mu\dagger} = - \gamma^{\mu}$, and $||F||^2 = \int d^4 x \,
F^2_{\mu\nu}(x)$. The last term is the second-order charge
renormalization subtraction at zero momentum transfer required for the
integral to converge for small $t$ in $d = 4$. A local counterterm is
not required to define $lndet$ in $d < 4$, and so the last term in
(\ref{Eq_Schwingerdet}) will be omitted in this case.

\subsection{$\mathbf{d = 2}$}

\label{Subsec_d=2}

\noindent
The calculation of (\ref{Eq_Schwingerdet}) in this case requires
knowledge of the bound and scattering states of a charged fermion
confined to a plane in the presence of a static magnetic field B
perpendicular to the plane. Of course there is actually no third
dimension in this strictly two dimensional problem. For
$A_{\mu} \in \bigcap_{n>2} L^{n}(\mathbb{R}^2)$,

\begin{equation}
\label{Eq_lndetStaticPerpB}
- - \frac{||e B||^2}{4 \pi m^2} \leq \text{ln det} \leq 0.
\end{equation}

\noindent
In addition, the lower bound requires that $B$ be square-integrable.
This bound will be derived in Sec. \ref{Sec_LowerBound}. A sharper
lower bound for a unidirectional magnetic field $B(x) \geq 0$ is given
by~\cite{Fry96a}

\begin{equation}
\label{Eq_LowerBoundUnidirectB}
\text{ln det} \geq \frac{1}{4 \pi} \int d^2 x \,
	\left[ e B(x) - [m^2 + e B(x)] \text{ln}
		\left( 1 + \frac{e B(x)}{m^2} \right) \right].
\end{equation}

\noindent
The upper bound in (\ref{Eq_lndetStaticPerpB}) was obtained as a
limit~\cite{Seiler82, Seiler81} of lattice
estimates.~\cite{Brydges79, Weingarten80} It is referred to
as the "diamagnetic" bound though physically it is a reflection of the
paramagnetism of a charged fermion in an external magnetic field as
(\ref{Eq_Schwingerdet}) makes clear. Moreover, the upper bound also
holds for nonabelian fields.~\cite{Seiler82, Seiler81, Brydges79} There
is the technical problem of proving that the lattice limit coincides
with the renormalized determinant $det_{ren}$ defined in
Refs. ~\citen{Seiler82, Seiler81, Weingarten79}
and that $det_{ren}$ can in turn be identified with definition
(\ref{Eq_Schwingerdet}). This is dealt with in
Refs.~\citen{Seiler82, Seiler81}. It would be desirable to
have a continuum proof of the upper bound in
(\ref{Eq_lndetStaticPerpB}).

If in addition $B$ has a finite range then, for $m^2 \to 0$

\begin{equation}
\label{Eq_FiniteRangeB}
\text{ln det} = \frac{|e \Phi|}{4 \pi} \, \text{ln} m^2 + R(m^2),
\end{equation}

\noindent
where $\lim_{m \to 0}[R(m^2)/\text{ln} \, m^2] = 0$ and
$\Phi = \int d^2 x \, B(x)$.~\cite{Fry00a} For integer values of
$e\Phi/2 \pi$ (\ref{Eq_FiniteRangeB}) is intuitively expected by the
Aharonov-Casher theorem~\cite{Aharonov79} which states that the number
of zero modes of $\slashletter{P} - e \slashletter{A}$ is
$[|e\Phi|/2 \pi]$, all with positive or negative chirality, where $[x]$
stands for the nearest integer less than $x$ and $[0] = 0$. This
result indicates that the zero-mass limit of $det$ does not coincide
with the Schwinger model except when $\Phi = 0$. The proof of
continuity at $m = 0$ will be outlined in Sec. \ref{Sec_ChiralLim}.

As a corollary of (\ref{Eq_FiniteRangeB}) suppose we define
$ln \, det_{3}$ as

\begin{equation}
\label{Eq_lndet3}
\text{ln det} = - \frac{e^2}{2 \pi}
	\int \frac{d^2 k}{(2 \pi)^2} \, | \hat{B}(k)|^2 \,
	\int^{1}_{0} dz \, \frac{z (1 - z)}{k^2 z(1 - z) + m^2} +
		\text{ln det}_3,
\end{equation}

\noindent
where the first term is just second-order perturbation theory and
$ln \, det_3$ is the remainder. Then provided $0 < |e \Phi| < 2\pi$,

there is at least one value of $m^2 > 0$ for which $ln \, det_3 =
0$.~\cite{Fry00b} This means the following: it is known that
$lndet_3(m^2 = 0) = 0$ as first shown by Schwinger.~\cite{Schwinger62}
Continuity of $ln \, det_3$ at $m^2 = 0$ when $\Phi = 0$ implies
$\lim_{m^2 = 0} ln \, det_3 = 0$. Equations (\ref{Eq_FiniteRangeB})
and (\ref{Eq_lndet3}) imply that $lndet_3 < 0$ if $0 < |e \Phi| < 2 \pi$
and $m^2$ is sufficiently small, while it must become positive before
approaching zero as $m^2 \to \infty$~\cite{Fry00b}. So our result
states that when $0 < |e \Phi| < 2 \pi$, the zero in $ln \, det_3$
moves from $m = 0$ when $\Phi = 0$ to some finite value(s) when
$m^2 > 0$. No information is available yet on mass zero(s) when
$|e \Phi| \geq 2 \pi$.

\subsection{$\mathbf{d = 3}$}

\label{Subsec_d=3}

\noindent
Here we have

\begin{equation}
\label{Eq_Boundsd=3}
- - \frac{Z}{6 \pi} \int d^2 x \, |e B(x)|^{\frac{3}{2}} \leq
	\text{ln det} \leq 0.
\end{equation}

\noindent
As for $d=2$ the upper bound was obtained as a limit
~\cite{Seiler82, Seiler81} of lattice estimates
~\cite{Brydges79} and holds also for nonabelian fields provided
$A_{\mu} \in \bigcap_{n > 3} L^n(\mathbb{R}^3)$. This restriction on $A_{\mu}$
is sufficient to make mathematical sense out of the renormalized
determinant $det_{ren}$ referred to above. The same technical problem
of limits referred to for $d = 2$ persists here.

The lower bound~\cite{Fry96b} is valid only for unidirectional fields
$B(x) \geq 0$ with $x \in \mathbb{R}^{2}$; hence the box cutoff $Z$ in the
direction of $B$.  In addition, the derivation assumed that
$A_\mu \in \bigcap_{n > 2} L^{n}(\mathbb{R}^2)$, that $B$ has finite flux and
$B \in L^{n}(\mathbb{R}^2)$, $n = 2, \frac{3}{2}$. A $2 \text{x} 2$
representation of the Dirac matrices was used. The definition (and
regularization) of the fermion determinant in (\ref{Eq_Schwingerdet})
is parity conserving and gives no Chern-Simons term, which is known to
be regularization dependent.~\cite{Deser91} The analysis leading to
the upper bound in (\ref{Eq_Boundsd=3}) predated the discovery of the
Chern-Simons term and is not included in the definition of $det$.

Another lower bound may be obtained from the lower bound in
(\ref{Eq_lndetStaticPerpB}) and the connection between the
$\text{QED}_2$ and $\text{QED}_3$ determinants,~\cite{Fry96b} namely

\begin{equation}
\label{Eq_RelationQEDdets}
\text{ln det}_{\text{QED3}} =
	\frac{Z}{2 \pi} \int^{\infty}_{m^2}
	\frac{d M^2}{\sqrt{M^2 - m^2}}
	\text{ln det}_{\text{QED2}}(M^2).
\end{equation}

\noindent
Eqs. (\ref{Eq_lndetStaticPerpB}) and (\ref{Eq_RelationQEDdets})
immediately give

\begin{equation}
\label{Eq_lndetQED3}
\text{ln det}_{\text{QED3}} \geq - \frac{Z ||e B||^2}{16 \pi |m|},
\end{equation}

\noindent
where in this bound $B$ no longer has to be unidirectional.

\subsection{$\mathbf{d = 4}$}

\label{Subsec_d=4}

\noindent
In this case only strong coupling limits are known. Thus if
$e \to \lambda e$, then

\begin{equation}
\label{Eq_Boundsd=4}
\frac{||e B||^2 Z T}{48 \pi^2} \leq
	\lim_{\lambda \to \infty}
	\left( \frac{\text{ln det}}{\lambda^2 ln \lambda} \right)
	\leq
	\frac{||e B||^2 \, T}{6 \pi^2}.
\end{equation}

\noindent
The upper bound~\cite{Fry96b} requires $A \in L^{3}(\mathbb{R}^{3})$ and
$B \in L^2(\mathbb{R}^3)$. If $\mathbf{A}$ is also in the Coulomb gauge
$\nabla \cdot \mathbf{A} = 0$ then we showed~\cite{Fry96b} that
$A_{\mu} \in \bigcap_{3 \leq n \leq 6} L^{n}(\mathbb{R}^3)$, which guarantees
that $ln \, det$ in (\ref{Eq_Boundsd=4}) is well-defined. Then the upper
bound in (\ref{Eq_Boundsd=4}) is valid for general square-integrable
static magnetic fields; $T$ is the (euclidean) time cutoff. The main
input to this bound is the upper bound on $ln \, det$ in $d = 3$ given by
(\ref{Eq_Boundsd=3}). The lower bound~\cite{Fry96a} holds only for
square-integrable magnetic fields $B(x) \geq 0$ and is derived from
the lower bound (\ref{Eq_LowerBoundUnidirectB}).

Finally we note that it is possible to make a derivative expansion of
$ln \, det$ into a sum of terms containing increasing numbers of
derivatives of $F_{\mu\nu}$.~\cite{Aitchison99} Such an approach has
the merit of going nonperturbatively beyond the constant field
approximation. Nevertheless, it is an expansion, and either all terms
are summed or the series is terminated and the remainder is
bounded. Investigation of this problem has begun with some exactly
solvable cases.~\cite{Dunne99} For the special case of a
$\text{sech}^2(x / \lambda)$ magnetic field it is found that the
derivative expansion of $ln \, det$ is a divergent asymptotic series
that is Borel summable.~\cite{Dunne99} This is encouraging. As far as
bounds on $ln \, det$ are concerned, all that is required is to show
that the series is asymptotic in the strict mathematical sense because
then one has a bound on the remainder of the series after any number
of terms.

\section{Lower Bound on ln det in $d = 2$}

\label{Sec_LowerBound}

\noindent
The lower bound in (\ref{Eq_lndetStaticPerpB}) was originally derived
for a unidirectional field $B \geq 0$ in
Ref. ~\citen{Fry96a}. We have since noticed that it is
easy to generalize this result to a general square-integrable
field. From here on the coupling constant e is absorbed into the
potential: $eA_{\mu} \to A_{\mu}$.

From definition (\ref{Eq_Schwingerdet}), without the charge renormalization term,

\begin{equation}
\label{Eq_lndetDeriv}
\pdo{m^2} \, \text{lndet} = \half
	\text{Tr}\left[(D^2 - \sigma_3 B + m^2)^{-1} - (P^2 + m^2)^{-1}\right],
\end{equation}

\noindent
where $D = P - A$. Using

\begin{multline}
\label{Eq_DerivT1}
(D^2 - \sigma_3 B + m^2)^{-1} = (D^2 + m^2)^{-1} +
	(D^2 + m^2)^{-1} \sigma_3 B (D^2 + m^2)^{-1}\\
	+ (D^2 + m^2)^{-1} \sigma_3 B (D^2 + m^2)^{-1}
	\sigma_3 B(D^2 - \sigma_3 B + m^2)^{-1},
\end{multline}

\noindent
(\ref{Eq_lndetDeriv}) becomes

\begin{multline}
\label{Eq_lndetDerivSub}
\pdo{m^2} \text{lndet} = \half \text{Tr} [ (D^2 + m^2)^{-1} - (P^2 + m^2)^{-1}\\
	+ (D^2 + m^2)^{-1}
	\sigma_3 B(D^2 + m^2)^{-1} \sigma_3
	B (D^2 - \sigma_3 B + m^2)^{-1} ].
\end{multline}

\noindent
Because of the definition (\ref{Eq_Schwingerdet}) the trace of the
first two terms in (\ref{Eq_lndetDerivSub}) is defined by the left-hand side of

\begin{equation}
\label{Eq_IntgTrace}
\int^{\infty}_0 dt \, \text{Tr} \left[ e^{-(P - A)^2 t} - e^{-P^2 t} \right]
	e^{-t m^2} \leq 0;
\end{equation}

\noindent
the right-hand side follows from Kato's inequality in the form stated by
the authors in Ref. \citen{Hogreve78}. Hence,

\begin{equation}
\label{Eq_LowerBoundDeriv}
\pdo{m^2} \text{lndet} \leq \half \text{Tr}
	\left[(D^2 + m^2)^{-1} \sigma_3 B (D^2 + m^2)^{-1} \sigma_3 B 
	(D^2 - \sigma_3 B + m^2)^{-1} \right].
\end{equation}

\noindent
By the Schwarz inequality, $|\text{Tr}(AB)| \leq ||A||_2 ||B||_2$,

\begin{equation}
\label{Eq_UseSchwarz}
\left|\text{Tr}[(D^2 + m^2)^{-1} \sigma_3 B (D^2 + m^2)^{-1}
	\sigma_3 B (D^2 - \sigma_3 B + m^2)^{-1}]\right|
	\leq ||\alpha||_2 ||\beta||_2,
\end{equation}

\noindent
where

\begin{eqnarray}
\nonumber
\alpha &=& (D^2 + m^2)^{-1} \sigma_3 B	\\
\label{Eq_AlphaBeta}
\beta  &=& (D^2 + m^2)^{-1} \sigma_3 B (D^2 - \sigma_3 B + m^2)^{-1}.
\end{eqnarray}

Setting  $<x|(D^2 + m^2)^{-1}|y> = \Delta(x,y | A)$,

\begin{eqnarray}
\nonumber
||\alpha||_2^2 &=& \text{Tr}( \sigma_3 B \Delta^{\dagger}(A) \Delta(A) \sigma_3 B)\\
\nonumber
               &\leq& \text{Tr}(B \Delta^2(A = 0) B)	\\
\label{Eq_Rearrange}
               &=& \frac{||B||^2}{2 \pi m^2},
\end{eqnarray}

\noindent
where use was made of the diamagnetic (Kato's) inequality

\begin{equation}
\label{Eq_KatoInequality}
| \Delta(x, y | A)| \leq \Delta(x - y),
\end{equation}

\noindent
in the form given in Ref. \citen{Vafa84}.

Next,

\begin{eqnarray}
\nonumber
||\beta||_2 &\leq& ||(D^2 - \sigma_3 B + m^2)^{-1}||_{\infty}
	||(D^2 + m^2)^{-1} \sigma_3 B||_2	\\
\label{Eq_LowerBoundBeta}
            &=& \frac{||\alpha||_2}{m^2},
\end{eqnarray}

\noindent
since $D^2 - \sigma_3 B \geq 0$.

Combining (\ref{Eq_LowerBoundDeriv}), (\ref{Eq_UseSchwarz}),
(\ref{Eq_Rearrange}), and (\ref{Eq_LowerBoundBeta}) gives

\begin{equation}
\label{Eq_UpperBoundDeriv}
\pdo{m^2} \text{lndet} \leq \frac{||B||^2}{4 \pi m^4}.
\end{equation}

\noindent
Now integrate (\ref{Eq_UpperBoundDeriv}) from $m^2$ to $m^2 = \infty$
and set $det(m^2 = \infty) = 1$. This does not conflict with any
renormalization condition. It is physically reasonable since an
infinite-mass fermion cannot respond to an external magnetic field. Then

\begin{equation}
\label{Eq_UpperBoundlndet}
\text{lndet} \geq - \frac{||B||^2}{4 \pi m^2},
\end{equation}

\noindent
which is the left-hand side of (\ref{Eq_lndetStaticPerpB}).

\section{Continuity of lndet in $d = 2$ at $m = 0$ when $\Phi = 0$}

\label{Sec_ChiralLim}

\noindent
From (\ref{Eq_lndet3}) we want to show that

\begin{eqnarray}
\nonumber
\lim_{m=0} \text{lndet}
	&=& \text{lndet}_{\text{Schwinger model}} \\
\label{Eq_lndetLimit}
	&=& \frac{1}{4 \pi^2} \int d^2x d^2y \,
		B(x) B(y) \,\, \text{ln}|x-y|,
\end{eqnarray}

\noindent
and in particular that

\begin{equation}
\label{Eq_lndet3Limit}
\lim_{m=0} \text{lndet}_3 = 0,
\end{equation}

\noindent
if $\Phi = 0$ and $A_{\mu} \in \bigcap_{n>1} L^n(\mathbb{R}^2)$. The analysis is
simplified if $B$ has finite range, which we will assume. Then the
limit in Eq. (\ref{Eq_lndetLimit}) comes from the first term in Eq.
(\ref{Eq_lndet3}). The demonstration of (\ref{Eq_lndet3Limit}) is not
in the literature to our knowledge. It is surprisingly tedious and
only its outline will be given.

\subsection{Step 1.}

\noindent
Let $S = (\slashletter{p} + m)^{-1}$. Neither $S\slashletter{A}$ nor
$(S\slashletter{A})^2$ are trace class operators, while
$(S\slashletter{A})^3$ is. Therefore the identity
$\text{lndet}(1 + A) = \text{Tr ln}(1 + A)$ for trace class operators
has to be modified to the regularized Fredholm determinant
\cite{Seiler81,Weingarten79,Dunford63,Simon77,Simon79,Seiler75b}

\begin{equation}
\label{Eq_lndetFredholm}
\text{lndet}_3 (1 - S\slashletter{A}) = \text{Tr}
	\left[ \text{ln}(1 - S\slashletter{A}) + S\slashletter{A}
	+ \half(S \slashletter{A})^2\right].
\end{equation}

\noindent
The connection of $lndet_3$ to definition (\ref{Eq_Schwingerdet}) is given
in Ref. ~\citen{Fry00b}.  The effect of (\ref{Eq_lndetFredholm}) is to remove the
ambiguous second-order graph from the determinant; it is defined by
the second-order expansion of (\ref{Eq_Schwingerdet}), giving the
first term in (\ref{Eq_lndet3}). The operator $S\slashletter{A}$ is a
compact operator on $L^2 (\mathbb{R}^2, \sqrt{k^2 + m^2} d^2 k; \mathbb{C}^2)$ for
$A_{\mu} \in \bigcap_{n>2} L^n(\mathbb{R}^2)$, $[\mathcal{C}_{n} = \{A |
\,||A||^{n}_{n} \equiv \text{Tr}(A^{\dagger} A)^{\frac{n}{2}} <
\infty\}]$.~\cite{Seiler82, Seiler81, Seiler75a, Seiler75b}
Because $S\slashletter{A} \in \mathcal{C}_3$, one more loop (i.e one
more order in $\slashletter{A}$) can be subtracted from the
determinant to give~\cite{Simon77,Simon79}

\begin{equation}
\label{Eq_lndet4}
\text{lndet}_4(1 - S\slashletter{A}) = \text{lndet}_3
(1 - S\slashletter{A}) + \frac{1}{3} \text{Tr}(S\slashletter{A})^3.
\end{equation}

\noindent
By C-invariance $\text{Tr}(S\slashletter{A})^3 = 0$. At this point it
is useful to make a similarity transformation
~\cite{Seiler82, Seiler81} and consider
$\text{det}_4 (1 - K_{m})$, where

\begin{equation}
\label{Eq_Km}
K_m = \frac{m - \slashletter{p}}{(p^2 + m^2)^{\frac{3}{4}}}
\slashletter{A} \frac{1}{(p^2 + m^2)^{\frac{1}{4}}},
\end{equation}

\noindent
is a compact operator on two-component square-integrable functions
on $\mathbb{R}^2$.

\subsection{Step 2.}

\noindent
The main theorem required to show that $\lim_{m=0} \text{lndet}_4 = 0$ is
Simon's theorem 6.5 specialized to $\text{det}_4$:

\begin{multline}
\label{Eq_lndet4Simon}
\left| \text{det}_4(1 - K_m) - \text{det}_4(1 - K_0)\right| \\
	\leq ||K_{m} - K_0||_4 \,\, \text{exp}
	\left[\Gamma_4(||K_m||_4 + ||K_0||_4 + 1)^4 \right],
\end{multline}

\noindent
where $\frac{1}{4} \leq \Gamma_4 \leq \frac{3}{4}$.~\cite{Simon77}
Here $K_0$ is $K_m$  with $m = 0$, which is known to be in
$\mathcal{C}_{n}$, $n > 2$.~\cite{Seiler80}

Furthermore,~\cite{Seiler80}

\begin{equation}
\label{Eq_det4}
\text{det}_4(1 - K_0) = 1.
\end{equation}

\noindent
It is essential for the proof of this that $A_{\mu}$ is "nonwinding"
so that $A_{\mu} \in \bigcap_{n>1} L^2(\mathbb{R}^2)$. [This condition was not
stated in Refs.~\citen{Seiler82, Seiler80}.  But it is
necessary, and the proofs leading to (\ref{Eq_det4}) still go
through.] Therefore by (\ref{Eq_lndet4Simon}), (\ref{Eq_det4}) and
(\ref{Eq_lndet4}) the demonstration of (\ref{Eq_lndet3Limit}) requires
that $\lim_{m=0} ||K_m - K_0||_4 = 0$.

\subsection{Step 3.}

\noindent
The operator difference $K_{m} - K_0$  may be decomposed into a
sum of five terms so that

\begin{multline}
\label{Eq_OpDiff}
||K_m - K_0||_4 \leq \left|\left| \frac{\slashletter{p}}{|p|}
	\left( \frac{1}{|p|^{\half}} \slashletter{A}
		\frac{1}{(p^2 + m^2)^{\frac{1}{4}}} -
		\frac{|p|}{(p^2 + m^2)^{\frac{3}{4}}} \slashletter{A}
		\frac{1}{(p^2 + m^2)^{\frac{1}{4}}} \right) \right|\right|_4\\
	+ \left|\left| \frac{\slashletter{p}}{|p|} \frac{1}{|p|^{\half}}
		\slashletter{A} \left( \frac{1}{|p|^{\half}} -
		\frac{1}{(p^2 + m^2)^{\frac{1}{4}}} \right) \right|\right|_4\\
	+ m \left|\left| \frac{1}{(p^2 + m^2)^{\frac{3}{4}}} \slashletter{A}
		\frac{1}{(p^2 + m^2)^{\frac{1}{4}}} \right|\right|_4.
\end{multline}

\noindent
The unitary $\slashletter{p}/|p|$ is irrelevant to the norms
and will be dropped from here on. We now outline the $m = 0$ limit of
the first norm.

By H\"{o}lder's inequality, $||fg||_r \leq ||f||_p ||g||_q$ for
$p^{-1} + q^{-1} = r^{-1}$, and $p, q, r \geq 1$,

\begin{multline}
\label{Eq_HoldersIneq}
\left|\left| \frac{1}{|p|^{\half}} \slashletter{A}
	\frac{1}{(p^2 + m^2)^{\frac{1}{4}}} -
	\frac{|p|}{(p^2 + m^2)^{\frac{3}{4}}} \slashletter{A}
	\frac{1}{(p^2 + m^2)^{\frac{1}{4}}} \right|\right|_{q} \\
	\leq
	\left|\left| \left( \frac{1}{|p|^{\half}}
		- \frac{|p|}{(p^2 + m^2)^{\frac{3}{4}}} \right)
		|A|^{\half} \right|\right|_p \left|\left| |A|^{\half}
		\frac{1}{(p^2 + m^2)^{\frac{1}{4}}}\right|\right|_p,
\end{multline}

\noindent
where $p = 2q > 4$ and the unitary $\slashletter{A}/|A|$  has
been dropped from the second norm in (\ref{Eq_HoldersIneq}). Then

\begin{equation}
\label{Eq_K1}
K_1 = \left( \frac{1}{|p|^{\half}}
	- \frac{|p|}{(p^2 + m^2)^{\frac{3}{4}}} \right) |A|^{\half},
\end{equation}

\noindent
is a compact operator in $\mathcal{C}_{n}$, $n > 2$ by Lemma 2.1 of
Seiler and Simon~\cite{Seiler75b} on integral operators of
the form $f(p)g(x)$ and, by the same lemma as stated in
Ref. ~\citen{Simon79},

\begin{equation}
\label{Eq_K1Norm}
||K_1||_{C_n} \leq (2 \pi)^{- \frac{2}{n}}
	\left|\left| \frac{1}{|p|^{\half}} -
	\frac{|p|}{(p^2 + m^2)^{\frac{3}{4}}}\right|\right|_{L^n}
	\left|\left| |A|^{\half} \right|\right|_{L^n}.
\end{equation}

\noindent
Simple estimates give $\lim_{m = 0} ||K_1||_{4 - \epsilon}  = 0$ for
$0 < \epsilon < 2$, and since $||K_1||_{4 - \epsilon} \geq ||K_1||_{4 +
\epsilon}$ is a general property of $\mathcal{C}_{n}$ spaces,

\begin{equation}
\label{Eq_K1NormLimit}
\lim_{m=0} ||K_1||_{4 + \epsilon} = 0.
\end{equation}

Referring again to (\ref{Eq_HoldersIneq}), let

\begin{equation}
\label{Eq_K2}
K_2(m) = |A|^{\half} \frac{1}{(p^2 + m^2)^{\frac{1}{4}}},
\end{equation}

\noindent
where $K_2 \in \mathcal{C}_{n}$, $n > 4$ by Lemma 2.1 of
Ref. ~\citen{Seiler75b}. At $m = 0$,
$K_2(0) = |A|^{\half} |p|^{-\half}$ and by writing $K_2(0)$ as the sum

\begin{equation}
\nonumber
|A|^{\half} \frac{1}{(m + |p|)^{\half}} + |A|^{\half}
	\left( \frac{1}{|p|^{\half}} - \frac{1}{(m + |p|)^{\half}} \right),
\end{equation}

\noindent
each of the two operators belongs to $\mathcal{C}_n$ by Lemma 2.1, and
hence so does $K_2(0)$.~\cite{SeilerPriv} More simple estimates give

\begin{equation}
\lim_{m=0}||K_2(m)||_{4 + \epsilon} \leq ||K_2(0)||_{4 + \epsilon},
\end{equation}

\noindent
where $0 < \epsilon < 2$. This together with (\ref{Eq_K1NormLimit}) and
(\ref{Eq_HoldersIneq}) give
                           
\begin{equation}
\label{Eq_ZeroLimit}
\lim_{m=0} \left|\left| \frac{1}{|p|^{\half}} \slashletter{A}
	\frac{1}{(p^2 + m^2)^{\frac{1}{4}}} -
	\frac{|p|}{(p^2 + m^2)^{\frac{3}{4}}} \slashletter{A}
	\frac{1}{(p^2 + m^2)^{\frac{1}{4}}} \right|\right|_q = 0
\end{equation}

\noindent
for $q > 2$, and hence the first norm on the right-hand side of
(\ref{Eq_OpDiff}) vanishes in the $m = 0$ limit.

Repetition of analysis of this kind establishes that the remaining
two norms in (\ref{Eq_OpDiff}) also vanish at $m = 0$, demonstrating
(\ref{Eq_lndet3Limit}) when $\Phi = 0$.

\section{Further Work}

\noindent
A major barrier to a better understanding of fermion determinants
beyond two dimensions is the lack of criteria for counting zero
modes. In three dimensions nothing is known, and in four dimensions
only the difference of positive and negative chirality zero modes is
known for a given background field. It is sometimes assumed that they all
have one or the other chirality. A counterexample would settle this
matter. In three dimensions the discovery of a topological invariant that
counts zero modes would be of considerable importance to physics as
well as mathematics.


\end{document}